%% file: lgpaper_hepex.tex
\documentclass{elsart2}
\journal{Physics Letters {\bf B}}

\usepackage{epsfig}
\usepackage{amssymb}

\usepackage{graphicx}

\def\Journal#1#2#3#4{{#1}~{\bf #2} (#4), #3}

\def\NIMA{{\em Nucl.~Instrum.~Methods}~A}
\def\NIMB{{\em Nucl.~Instrum.~Methods}~B}

\def\PLB{{\em Phys.~Lett.}~B}
\def\PRL{\em Phys.~Rev.~Lett.}
\def\PRD{{\em Phys.~Rev.}~D}
\def\PR{\em Phys.~Rev.}
\def\EPJ{{\em Eur.~Phys.~J.}~C}
\def\JMPA{{\em Int.~J.~Mod.~Phys.}~A}
\newcommand{\subrm}[1]{\mbox{\tiny \rm #1}}
\newcommand{\Br}{\rm Br}
\newcommand{\xilg}{\Xi^0 \to \Lambda \gamma}

\newcommand{\XiLgam}{\xilg}
\newcommand{\xilpiz}{\Xi^0 \to \Lambda \pi^0}

\newcommand{\XiLpi}{\xilpiz}
\newcommand{\ctl}{\cos \Theta_\Lambda}
\newcommand{\Xiz}{{\Xi}^0}
\newcommand{\xiz}{{\Xi}^0}
\newcommand{\lppic}{\Lambda\to p \pi^-}
\newcommand{\Lamppi}{\Lambda\to p \pi^-}
\newcommand{\pim}{{\pi}^-}

\begin{document}



\begin{frontmatter}
\title{Measurement of the $\xilg$ Decay Asymmetry and Branching Fraction}
\date{}
\input{authorlist_hepex}
\vspace*{\fill}
\begin{abstract}
In data taken with the NA48 experiment at the CERN SPS in 1999,
730 candidates of the weak radiative hyperon decay $\xilg$ have been found with an estimated background of $58 \pm 8$ events.
From these events the $\XiLgam$ decay asymmetry has been determined to
$\alpha(\XiLgam) = -0.78 \pm 0.18_{\subrm{stat}} \pm 0.06_{\subrm{syst}}$,
which is the first evidence of a decay asymmetry in $\XiLgam$.
The branching fraction of the decay has been measured to be
$\Br(\XiLgam) = (1.16 \pm 0.05_{\subrm{stat}} \pm 0.06_{\subrm{syst}}) \times 10^{-3}$.
\end{abstract}

\end{frontmatter}

\setcounter{footnote}{0}

\setcounter{section}{1}                                        %
\renewcommand{\section}[1]{\vspace*{4mm}                       
                          {\bf \thesection \hspace*{2mm} #1}   
                          \addtocounter{section}{1}}           %
\newcommand{\spaceafterfloat}{\vspace*{3mm}} 

%
%

\section{Introduction}

The $\xilg$ decay asymmetry plays an important role in solving a long standing discrepancy
between the Hara theorem \cite{bib:hara64} and the observed decay asymmetries of
weak radiative hyperon decays \cite{bib:zenc96,bib:zenc99-4}.
The Hara theorem states that the parity-violating amplitude of weak radiative hyperon decays vanishes in the SU(3) limit.
Accordingly, the decay asymmetries will vanish in this case.
Introducing weak breaking of SU(3) symmetry one expects to observe small decay asymmetries.
In contrast to this, a large negative decay asymmetry in the weak radiative decay $\Sigma^+\to p\gamma$ 
was first measured by Gershwin {\it et al.}~\cite{bib:gershwin69}
and later confirmed~\cite{bib:pdg02}.
To address this observation, models were developed which tried to obtain large decay asymmetries
in spite of weak SU(3) breaking. 
One category consists of pole models, which satisfy the Hara theorem by construction,
and approaches based on chiral perturbation theory.
They predict negative decay asymmetries for all weak radiative hyperon decays~\cite{bib:gavela81,bib:borosay99,bib:zenc00}.
Vector meson dominance models and calculations based on the quark model on the other hand violate the Hara theorem.
It has been shown that the Hara theorem is generally violated in quark model approaches \cite{bib:kamal83}.
This second group of models favours a positive decay asymmetry for the channel $\XiLgam$~\cite{bib:verma88,bib:lach95}.
Therefore, the decay $\XiLgam$ plays a crucial role
in differentiating between the groups of models.
The only previous measurement of the $\xilg$ decay asymmetry
has been performed at FNAL, giving a value of $-0.43 \pm 0.44$~\cite{bib:james90}\footnote{
The original publication~\cite{bib:james90} quoted a wrong sign of the asymmetry.
This has been corrected in an erratum in 2002, but not yet in the PDG value~\cite{bib:pdg02}.},
and is not able to make the distinction.

In addition to the decay asymmetry, the branching fractions of radiative $\Xi^0$ decays
are similarly difficult to calculate.
Predictions on the $\XiLgam$ decay rate vary by about one order 
of magnitude~\cite{bib:gavela81,bib:borosay99,bib:zenc00,bib:verma88,bib:gilman79,bib:kamal82}.
Two experiments have reported a branching fraction measurement~\cite{bib:james90,bib:na4800},
with the measured values being about two standard deviations apart.
The latter measurement is an earlier NA48 measurement based on a data set of the year 1997,
which contained only 31 $\XiLgam$ candidates.

In this letter we report a new measurement with greatly enhanced precision
on both the $\XiLgam$ decay asymmetry and branching fraction.

\section{Experimental setup}

The NA48 experiment has been designed for the measurement of direct CP violation
in the decay of neutral kaons. However, due to the comparable life times,
decays from $\Lambda$ and $\Xi^0$ hyperons could also be recorded by the NA48 detector.
The experiment featured two almost collinear beams of neutral kaons which were derived from 450~GeV/$c$ protons 
hitting two fixed beryllium targets of 40~cm length and 0.2~cm diameter,
which were located 6~m (near target) and 126~m (far target), respectively, upstream of the decay region~\cite{bib:na48_beams}.
The present measurement is based on data from a special two-day run period in 1999
where only the near target was in operation.
The proton beam was extracted from the CERN~SPS in a 2.4~s spill every 14.4~s with the intensity increased
to about $5\times10^9$ protons per spill, which is a factor of $\sim 200$ higher than in normal running conditions.
The near target was located 7.2~cm above the primary proton beam axis, which points to the detector centre.
It was followed by a sweeping magnet with tungsten-alloy inserts in which the protons
not interacting in the target were absorbed.
The neutral beam was defined by a 0.36~cm diameter collimator 480~cm downstream of the target.
It had a production angle of 4.2~mrad and a downward inclination of 0.6~mrad with respect to the
incident proton beam, in order to have the axis of the neutral beam pointing to the centre of the electromagnetic calorimeter.  
The decay region was contained in an evacuated 89~m long and 2.4~m diameter steel tank,
closed by a $0.3\%$ radiation lengths thick polyamide (Kevlar) composite window, from the centre of which
a beam pipe traverses the following main NA48 detector.

The main NA48 detector elements are a magnetic spectrometer and a liquid-krypton electromagnetic calorimeter.
The spectrometer is composed of two drift chambers~\cite{bib:dch} upstream and two downstream of a dipole magnet.
The magnetic field is directed vertically and produces a 265~MeV/$c$ transverse momentum kick.
The chambers provide 100~$\mu$m spatial resolution on the track coordinates,
resulting in a total momentum resolution of $\sigma(p)/p =0.6\%$ for 45~GeV/$c$ particles.

The liquid-krypton calorimeter (LKr) measures the energies, positions, and times
of electromagnetic showers initiated by photons and electrons.
Around 20~t of liquid krypton at 121~K are used in a ionization detector. The calorimeter
has a structure of 13212 square read-out cells of $2\times 2$~cm$^2$ cross-section and 127~cm length
(27 radiation lengths) each. The cross-section of the active volume has an octagonal shape
of 240~cm inscribed diameter. 
The read-out cells are formed by longitudinally stretched copper-beryllium ribbons which act as electrodes.
The energy resolution of the calorimeter is 
$\sigma(E)/E = 9\% / E \oplus 3.2\% / \sqrt{E} \oplus 0.42\%$ with $E$ in GeV; the time resolution
is better than 300~ps for photon energies above 20~GeV.

Further detector elements were used in the trigger decision for the present measurement: 
a steel-scintillator sandwich calorimeter (HAC) with a length of 6.7 nuclear interaction lengths follows the
liquid-krypton calorimeter and measures energies and horizontal and vertical positions of hadron showers; 
a hodoscope composed of 64 horizontal and 64 vertical scintillator strips 
is located about 1.6~m upstream of the LKr calorimeter.
A more detailed description of the experimental set-up can be found elsewhere \cite{bib:lai01}.

\begin{figure}[t]
\begin{center}
\epsfig{file=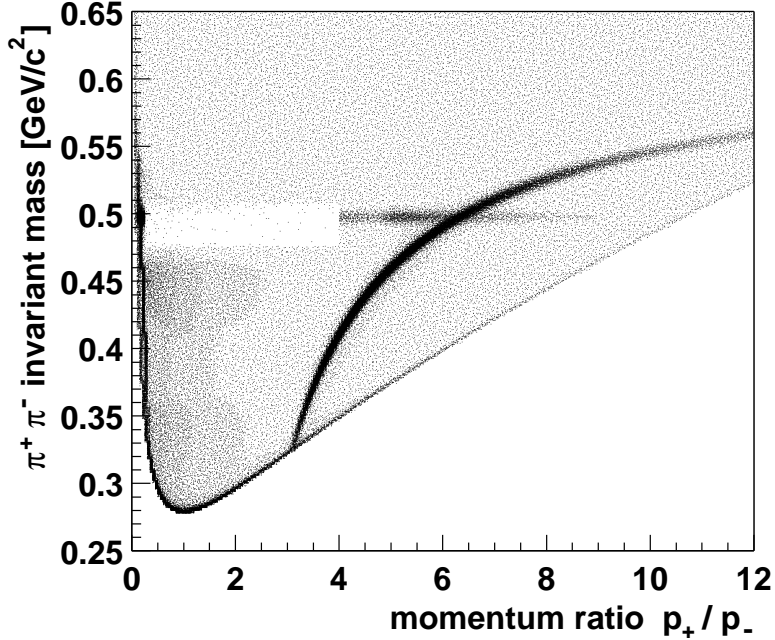,width=0.68\textwidth}
\caption{Two-track invariant mass with $\pi^+ \pi^-$ assumption versus
the momentum ratio of positive to negative tracks.
Only trigger cuts are applied. The curved structure starting at $p_+/p_- \simeq 3$ consists of
$\Lambda \to p \pi^-$ decays.}
\label{fig:banana}
\end{center}
\spaceafterfloat
\end{figure}

The trigger decision for neutral hyperon decays was based on information from the detector elements described above.
A positive pre-trigger decision required at least one coincidence between a vertical and a horizontal
hodoscope scintillator strip, a two track signature in the drift chambers, and
the sum of the energies deposited in the electromagnetic and the hadron calorimeter larger than 35~GeV.
The next trigger level used information from a preliminary track reconstruction:
In order to reject $K_S \to \pi^+ \pi^-$ background events, a mass cut around the kaon mass for the $\pi^+ \pi^-$ hypothesis 
and a cut on the ratio $p_+/p_-$ between the positive and negative track momenta were applied.
This ratio is large for $\lppic$ decays
where the proton carries the major fraction of the initial momentum, as opposed to background kaon decays
where the two charged particles typically have a much lower momentum ratio.
The trigger rejected a window with $1/5 < p_+/p_- < 5$ ($1/4$ and 4 for a brief period at the start of the run)
and $|m_{\pi^+ \pi^-} - m_K| < 30$~MeV/c$^2$ (25~MeV/c$^2$), where $m_K$ is the nominal kaon mass.
Because of the high rate the triggered events were down-scaled with different factors between 4 and 8
during the two-day run period.
The plot of $m_{\pi^+ \pi^-}$ versus $p_+/p_-$ for the accepted events is shown in Fig.~\ref{fig:banana}.
In total, approximately 150 million triggers were collected, out of which 18~million events
contained $\Lambda$ hyperons.

The events were selected by a trigger optimised to
select hadronic events for which the accuracy of the electromagnetic
calorimeter information was not necessary. For this trigger, showers
from the LKr were read-out only if their seed energy exceeded 1~GeV
instead of the usual 100~MeV.
Therefore a cluster energy correction was applied, which was determined from the
reconstructed invariant $\pi^0$ mass in $K_L \to \pi^+ \pi^- \pi^0$ events from the same data sample.
The correction was energy-dependent and ranged between $3\%$ and $0.5\%$ for photon energies between
5 and 20~GeV.

A Monte Carlo simulation based on GEANT~3.21~\cite{bib:geant} for the LKr calorimeter
and a fast track simulation was used throughout this analysis.
Special care was taken to adjust beam, target, and collimator geometries and positions to the data.

\section{Method of the Asymmetry Measurement}

\begin{figure}[t]
\begin{center}
\epsfig{file=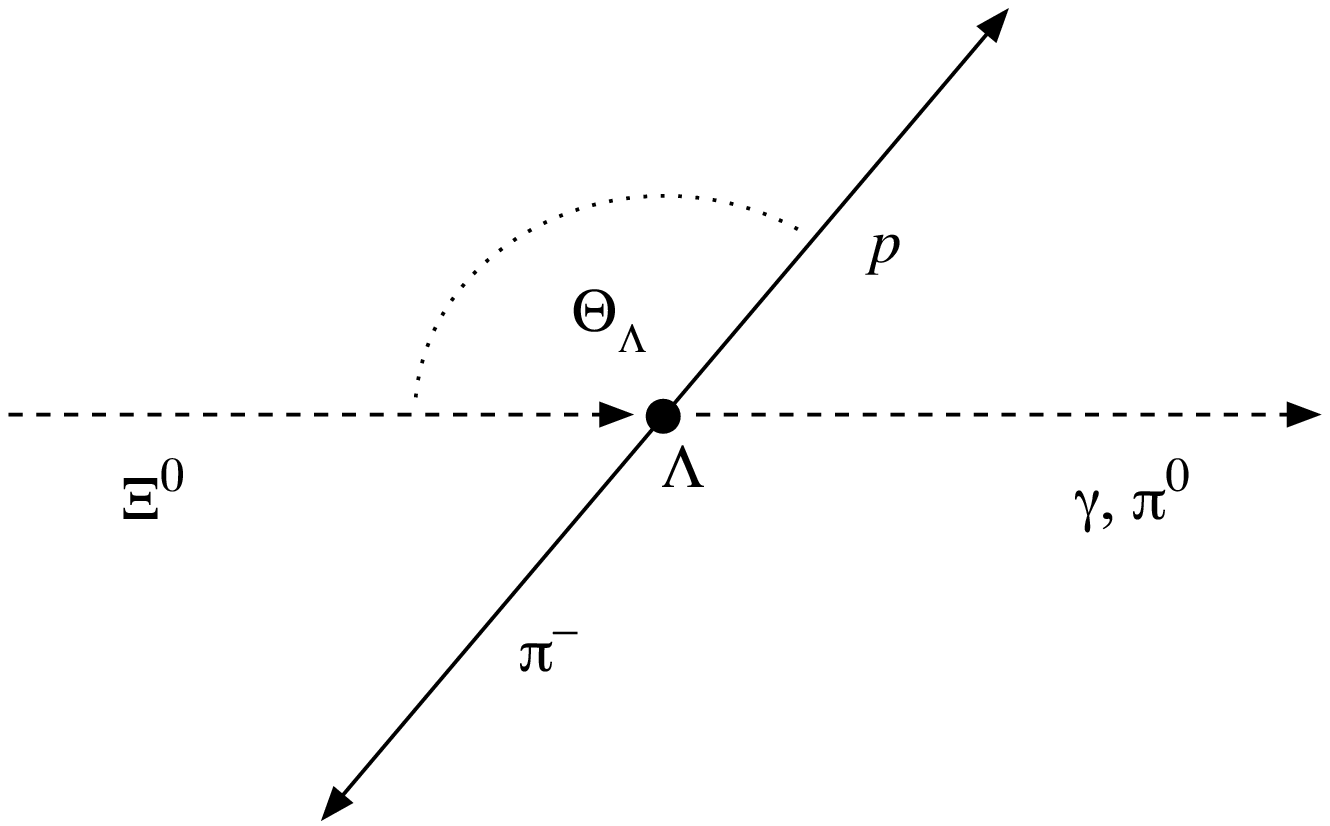,width=0.5\textwidth}
\caption{\label{fig:asym}The decay angle $\Theta_\Lambda$ in the $\Lambda$ rest frame,
as measured for both the $\xilg$ and the $\xilpiz$ decay.}
\end{center}
\spaceafterfloat
\end{figure}

For the asymmetry measurement we exploit the well-known decay asymmetry of the $\lppic$ decay.
The $\Lambda$ hyperons are longitudinally polarized by the parent process $\xilg$
with a polarization of $\alpha(\XiLgam)$ in their rest frame.
Effectively, one measures the distribution of the angle $\Theta_\Lambda$ between the incoming $\xiz$ (corresponding to
the outgoing $\Lambda$ direction in the $\xiz$ rest frame) and the outgoing proton
in the $\Lambda$ rest frame (see Figure \ref{fig:asym}):
\[
{dN \over d\cos\Theta_\Lambda} = N_0 \left( 1 - \alpha(\Lamppi) \, \alpha(\XiLgam) \, \cos \Theta_\Lambda \right)
\]
\vspace*{-5mm}
In this way the $\Lambda$ is polarized by the $\Xiz$ decay and analysed by its own decay into $p\pim$.
The minus sign is purely conventional and arises from the fact that the photon carries spin 1, which leads to an opposite
$\Lambda$ spin to that in the process $\XiLpi$~\cite{bib:behrends58}.

For calibration purposes and as a cross-check, we have also analyzed the decay $\XiLpi$,
for which the decay asymmetry is well measured.
Obviously, when replacing $\alpha(\XiLgam)$ with $\alpha(\XiLpi)$,
there is no difference in the topology for the decays $\XiLgam$ and $\XiLpi$
and the definition of the angle $\Theta_\Lambda$ is similar.
However, as explained above, the spin~0 nature of the $\pi^0$ 
leads to a sign flip for the longitudinal $\Lambda$ polarization.

\section{Event Selection}

The $\Lambda$ hyperons were identified by two oppositely charged tracks in the spectrometer.
The tracks were required to be at most 4.5~ns apart in time and
to have momenta of at least 30~GeV/$c$ for the positive and 5~GeV/$c$ 
for the negative track.
For both tracks the radial distances to the beam axis at the longitudinal positions of
the drift chambers had to be greater than 12.6~cm to ensure full efficiency
of the chambers.
The extrapolation of the two tracks 
to the point of closest approach defined the $\Lambda$ decay vertex. The 
distance of closest approach had to be less than 2.2~cm.
The reconstructed $\Lambda$ hyperon was required to have a minimum momentum of 57~GeV/$c$ and
a reconstructed invariant mass within
2.7~MeV/$c^2$ of the nominal $\Lambda$ mass, corresponding to a window of $\pm 3$~standard deviations.
Finally, the longitudinal position of the $\Lambda$ decay vertex had to be at least 6~m downstream of the target,
corresponding to the end of the final collimator, and at most 40~m from the target.

Photons were detected by clusters in the electromagnetic calorimeter.
The cluster energy had to be above 15~GeV.
To ensure negligible energy loss, all clusters had to be at least 15~cm and at most 120~cm
from the beam axis, well within the sensitive volume of the LKr calorimeter,
and had to have a distance greater than 2~cm to any dead calorimeter cell.
To avoid effects from hadron shower contaminations,
the photon clusters were required to be separated by at least 25~cm from the impact point of any track.
The cluster-time had to be within $\pm 3$~ns of the average of the track times.

The energy centre-of-gravity at the longitudinal position of the calorimeter
was defined as
\begin{displaymath}
\vec{r}_{cog} = \sum_i \vec{r}_i E_i \: / \: \sum_i E_i, 
\end{displaymath}
\vspace*{-5mm}
with the sums running over tracks and photon clusters. The positions $\vec{r}_i$
are either the cluster positions in case of photons or the coordinates
of the tracks extrapolated from before the spectrometer to the LKr calorimeter.
The energies $E_i$ are the photon cluster energies or are derived from the track momenta
using the nominal particle masses.
To avoid missing energy in the event,
the position of the energy centre-of-gravity was required to be less than 7~cm from the detector axis.
The target position and the energy centre-of-gravity $\vec{r}_{cog}$
defined the reconstructed line-of-flight of the $\Xi^0$.
The distance of the $\Lambda$ decay vertex from the $\Xi^0$ line-of-flight
had to be less than 2~cm to ensure a well-reconstructed event.
The $\Xi^0$ decay vertex was reconstructed as the point of closest approach between the 
$\Xi^0$ line-of-flight and and the $\Lambda$ line-of-flight, which was determined from the
reconstructed $\Lambda$ decay vertex and 3-momentum.
The longitudinal position of the $\Xi^0$ decay vertex was required to be at least 5~m
downstream of the target, 1~m before the end of the final collimator.

Background from $K_S \to \pi^+ \pi^- \gamma$ decays was completely discarded by rejecting all events
with an invariant mass, under a $\pi^+ \pi^- \gamma$ hypothesis, within $7.5$~MeV/$c^2$
of the nominal kaon mass (corresponding to $2.5$~standard deviations of the resolution).
Additional background comes from $K_L \to \pi^+ \pi^- \pi^0$ decays as well as
from time overlap of two events or hadronic interactions in the collimator region.
Both accidental and hadronic background contain a real $\Lambda$ decay
together with one or two photons from a $\pi^0$ decay.
In order to reject this background together with $K_L \to \pi^+ \pi^- \pi^0$ decays,
we rejected all events which had one or more additional LKr calorimeter cluster
within 5~ns of the average track-time and more than 25~cm from the track impact points
in the calorimeter.
Decays of $\XiLpi$ with one lost photon are strongly suppressed by kinematics 
to enter the signal region, even if resolution effects are taken into account.

\begin{figure}[t]
\begin{center}
\epsfig{file=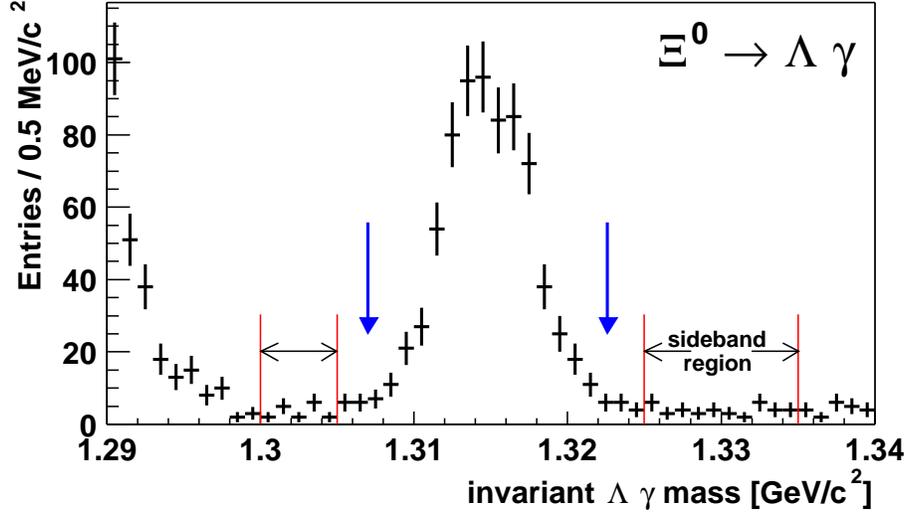,width=0.8\textwidth}
\caption{Distribution of the invariant $\xiz$ mass for $\xilg$ decays.
The signal region is indicated by the vertical arrows.
Also shown are the side-band regions used for the background subtraction.
The sharp increase of events below $1.3$~GeV/$c^2$ is due to
$\XiLpi$ events with a lost photon.}
\label{fig:lg_inv_mass}
\end{center}
\spaceafterfloat
\end{figure}

The $\Lambda \gamma$ invariant mass distribution of the $\Xi^0$ candidates
is shown in Figure~\ref{fig:lg_inv_mass}.
Within $\pm 7.8$~MeV/$c^2$ (3 standard deviations) of the $\xiz$ mass 730 $\xilg$ candidates
were reconstructed.
As all remaining background sources have a flat distribution in a close invariant $\Lambda \gamma$ mass interval,
as seen from Monte Carlo simulation,
we have used mass side-bands from
$1.3$ to $1.305$~GeV/$c^2$ and from $1.325$ to $1.335$~GeV/$c^2$ for background subtraction.
From this, the background was estimated to be $58.2 \pm 7.8$ events, 
which corresponds to $(8.0 \pm 1.1)\%$ of the signal sample.
It is dominated by hadronic events with some remaining contributions from
accidental background and mis-identified $K_L \to \pi^+ \pi^- \pi^0$ decays.

For the selection of the normalization channel $\XiLpi$ only a few changes 
were applied with respect to the $\XiLgam$ selection.
At least two photon clusters within 5~ns and with energies above 6~GeV were required.
The invariant mass of the two photons, computed using the cluster energies and positions and 
the reconstructed $\Xi^0$ decay vertex, had to be within 3 standard deviations (about 10~MeV/$c^2$) 
from the nominal $\pi^0$ mass. 
No cut on additional LKr calorimeter clusters was applied for the $\XiLpi$ selection.

\begin{figure}[t]
\begin{center}
\epsfig{file=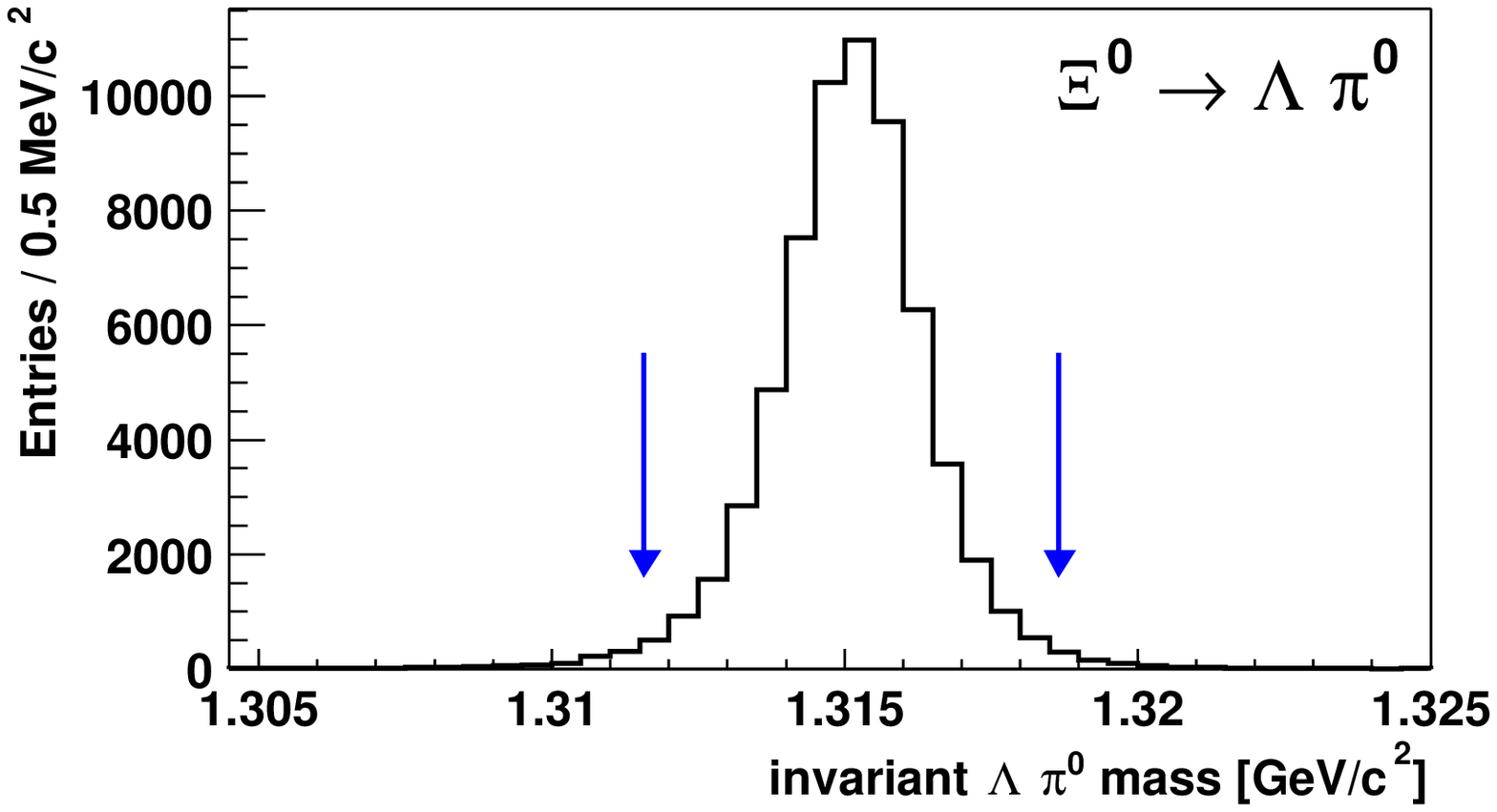,width=0.8\textwidth}
\caption{Distribution of the invariant $\xiz$ mass for reconstructed $\xilpiz$ decays.
The signal region is indicated by the arrows.}
\label{fig:lpi0_inv_mass}
\end{center}
\spaceafterfloat
\end{figure}

After applying all selection criteria, the total yield of $\XiLpi$ candidates was 61828 events
within a window of $\pm 3.0$~MeV/$c^2$ ($3\sigma$) around the nominal $\Xi^0$ mass (Fig.~\ref{fig:lpi0_inv_mass}).
Backgrounds are negligible in this channel.

\section{Measurement of the Decay Asymmetry}

\begin{figure}[t]
\begin{center}
\epsfig{file=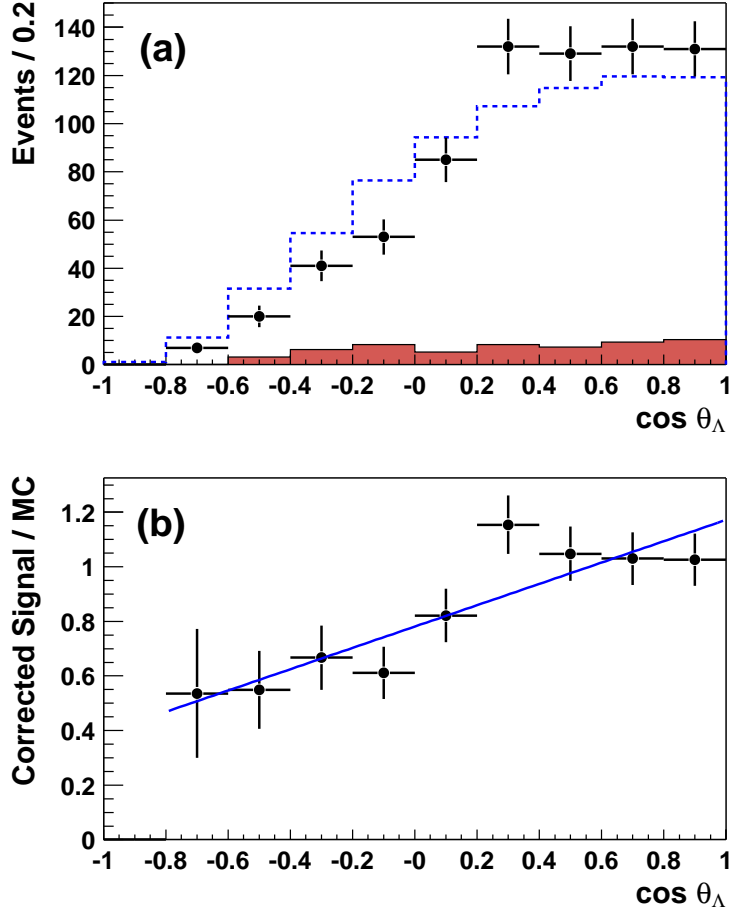,width=0.75\textwidth}
\caption{The $\XiLgam$ decay asymmetry:
(a) $\ctl$ distributions of signal candidates (crosses), scaled side-band events (shaded),
    and isotropic Monte Carlo events (dashed).
(b) Ratio of background-corrected signal candidates over isotropic Monte Carlo simulation.
    The line shows the result of the fit.}
\label{fig:lg_asym}
\end{center}
\spaceafterfloat
\end{figure}

As described, the asymmetry is analysed in the angle $\ctl$. 
The resolution on $\ctl$ is 0.01, as determined from the Monte Carlo
simulation.
Fig.~\ref{fig:lg_asym}a shows
the $\ctl$ distributions for all $\XiLgam$ candidates
and for the properly scaled side-band events
compared to the isotropic Monte Carlo simulation.

The ratio of data over Monte Carlo corrects for the detector acceptance.
To account for the background contamination, 
an effective asymmetry $\alpha_{\subrm{bkg}}$ of the background was determined by fitting
the side-band events with $dN/d\ctl \propto 1 - \alpha_{\subrm{bkg}} \ctl$.
The result of $\alpha_{\subrm{bkg}} = 0.13 \pm 0.32$ (together with the background fraction of $(8.0\pm1.1)\%$)
was used to apply a smooth correction to the $\ctl$ distribution of the signal events for the background contamination.
The ratio of background-corrected signal over Monte Carlo is shown in Figure~\ref{fig:lg_asym}b.
A fit with the normalization
and the asymmetry $\alpha(\Lamppi) \alpha(\XiLgam)$ as free parameters was performed in the range
$-0.8 < \ctl < 1$, where both data and Monte Carlo statistics are large.
The result of the fit is
$\alpha(\Lamppi) \alpha(\XiLgam) = -0.50 \pm 0.12_{\subrm{stat}}$
with a $\chi^2$ per degree of freedom of $9.4/7$.
After dividing by $\alpha(\Lamppi) = 0.642 \pm 0.013$~\cite{bib:pdg02} we obtain
$\alpha(\XiLgam) = -0.78 \pm 0.18_{\subrm{stat}}$.

Several contributions to the systematic error have been studied.
The largest uncertainties come from the effective asymmetry of the background ($\pm 0.052$)
and the background normalization  ($\pm 0.019$).
Additional uncertainties may arise from a possible incorrect modeling of the detector acceptance.
This was investigated with $\XiLpi$ events, which have sufficient statistics, and 
which are, due to the lower $Q$ value, more sensitive to possible inefficiencies near the beam pipe.
Different geometrical acceptance cuts in the beam pipe regions and on the $\Xi^0$ energy 
as well as uncertainties in the beam profile
simulation and the $p_T$ distribution of the $\Lambda$ hyperons have been studied,
resulting in a variation of $\pm 0.007$ on the decay asymmetry.
The effect of transverse $\Xi^0$ polarization on the decay asymmetry was  
studied and found to be negligible compared to the other uncertainties.
Finally, the uncertainty on the used value of $\alpha(\Lamppi)$ contributes $\pm 0.016$ to the systematics.

By adding the single components in quadrature, the total systematic 
uncertainty is $\pm 0.06$, which is small compared to
the statistical uncertainty of the data (Tab.~\ref{tab:asymsyst}).

An important check for the validity of the $\XiLgam$ analysis is the measurement of the decay asymmetry in $\XiLpi$, where
the data statistics is a factor of 100 higher. Both channels are two-body decays and the decay $\XiLpi$, due to the lower $Q$ value,
is even more sensitive to effects from the beam-line geometry and polarization than $\XiLgam$.
The decays $\XiLpi$ were recorded with the same trigger, analysed with the same definition of $\ctl$, and used a similar Monte Carlo simulation
as for the $\XiLgam$ analysis.
From our data we measure a combined $\XiLpi$ asymmetry of $\alpha(\Lamppi) \alpha(\XiLpi) = -0.257 \pm 0.011_{\subrm{stat}}$.
The excellent agreement between this result and the best published measurement of $\alpha(\Lamppi) \alpha(\XiLpi) = -0.260 \pm 0.006$~\cite{bib:handler82}
validates the method and serves as a systematic check.

As a further check, a second analysis, which was performed using the same data but different Monte Carlo samples,
gave consistent results.

\begin{table}[th]
\begin{center}
\begin{tabular}{lr}
\hline \hline
                                     & $\Delta \: \alpha(\XiLgam) $ \\ \hline
Background asymmetry                 &   $\pm \: 0.052$          \\
Background normalization             &   $\pm \: 0.019$          \\
Detector acceptance                  &   $\pm \: 0.007$          \\
$\alpha(\Lamppi) = 0.642 \pm 0.013$ &   $\pm \: 0.016$          \\ \hline 
Total systematic uncertainty         &   $\pm \: 0.06$          \\*[1mm]
Statistical uncertainty data         &   $\pm \: 0.18$          \\
\hline \hline
\end{tabular}
\caption{Summary of uncertainties on the $\XiLgam$ decay asymmetry $\alpha(\XiLgam)$.}
\label{tab:asymsyst}
\end{center}
\end{table}

\section{Measurement of the $\XiLgam$ Branching Fraction}

Using the central value of our new result on the decay asymmetry, the overall 
$\XiLgam$ reconstruction efficiency was determined
to be $5.4\%$ from Monte Carlo simulation. 
For the normalization channel $\XiLpi$ the reconstruction efficiency 
was determined to be $0.58\%$. It is significantly lower than for the signal channel,
as the proton receives less transverse momentum and escapes more often undetected through the beam-pipe.
Subtracting the background and using $\Br(\XiLpi) = 99.5\%$~\cite{bib:pdg02}, we obtain
$\Br(\XiLgam) = (1.16 \pm 0.05_{\subrm{stat}}) \times 10^{-3}$.

The systematic error is dominated by the uncertainty on the decay asymmetry,
which leads to a $\pm 4.8\%$ uncertainty on the geometrical acceptance and therefore on the branching ratio.
Other systematic uncertainties are the detector acceptance ($\pm 1.1\%$ of the measured value), the $\Xi^0$ polarization ($\pm 1.4\%$), 
and the background estimation ($\pm 1.1\%$), which have been determined in a similar way to those for
the decay asymmetry measurement.
To account for the unknown $\Xi^0$ polarization in our data we have varied the polarization
in the simulation by $\pm 10\%$, which is similar in magnitude to the
measured transverse $\Xi^0$ polarization of about $-10\%$
for a proton beam energy of 800~GeV and a production angle of 4.8~mrad~\cite{bib:ktevpol}.
This variation resulted in a $\pm 1.4\%$ change of the measured branching fraction.
An additional external uncertainty arises from the knowledge of the $\XiLpi$ decay asymmetry ($\pm 0.5\%$),
since it affects the selection efficiency of the normalization channel.
Finally, the statistics of the normalization channel and of the signal and normalization Monte Carlo simulations contribute
$\pm 0.7\%$ to the systematic uncertainty.
All other sources of systematic errors were found to be negligible.
The total systematic uncertainty on $\Br(\XiLgam)$ is $\pm 5.3\%$.
A summary of uncertainties on the branching ratio measurement is given in Table~\ref{tab:brsyst}.

A second independent analysis has been performed also for the branching 
fraction measurement and yielded a consistent result.

\begin{table}[t]
\begin{center}
\begin{tabular}{lr}
\hline \hline
                              & $\Delta \: \Br(\XiLgam)$  \\
                              & $[10^{-3}]$               \\ \hline
$\XiLgam$ decay asymmetry     &   $\pm \: 0.056$          \\
$\XiLpi$  decay asymmetry     &   $\pm \: 0.006$          \\
Background estimation         &   $\pm \: 0.013$          \\
Detector acceptance           &   $\pm \: 0.013$          \\
$\Xi^0$ polarization          &   $\pm \: 0.016$          \\ 
Statistics of MC and $\XiLpi$ &   $\pm \: 0.008$          \\ \hline
Total systematic uncertainty  &   $\pm \: 0.06$           \\*[1mm]
Statistical uncertainty data  &   $\pm \: 0.05$           \\
\hline \hline
\end{tabular}
\caption{Summary of uncertainties on $\Br(\XiLgam)$.}
\label{tab:brsyst}
\end{center}
\end{table}

\vspace*{-4mm}
\section{Summary and Conclusion}

From the data of two days of high intensity $K_S$ data taking of the NA48 experiment
730 $\xilg$ candidates have been found with an estimated background of $58.2 \pm 7.8$ events.
By comparison with an isotropic Monte Carlo simulation, the $\xilg$ decay asymmetry has been found to be
\[
\alpha(\XiLgam) =  -0.78 \pm 0.18_{\subrm{stat}} \pm 0.06_{\subrm{syst}}.
\]
\vspace*{-4mm}
This is the first evidence for a negative decay asymmetry in this channel~\cite{bib:sas}. 
As has been indicated in the introduction, the $\xilg$ decay asymmetry
is a crucial input parameter for theoretical approaches to weak radiative hyperon decays.
The negative sign of the asymmetry clearly prefers models consistent with the Hara theorem, while, 
on the other hand, it cannot be easily explained by quark models and models using vector meson dominance.

In addition we have determined the $\xilg$ branching fraction to
\[
\Br(\XiLgam) = (1.16 \pm 0.05_{\subrm{stat}} \pm 0.06_{\subrm{syst}}) \times 10^{-3},
\]
\vspace*{-4mm}
which is the most precise measurement of this decay rate so far.
It confirms the previous FNAL measurement~\cite{bib:james90}, while being about 1.9~standard
deviations below the previous NA48 measurement~\cite{bib:na4800}.


\section{Acknowledgements}

It is a pleasure to thank the technical staff of the participating
laboratories, universities, and affiliated computing centres for their
efforts in the construction of the NA48 apparatus, in the
operation of the experiment, and in the processing of the data.
We also would like to thank Piotr \.{Z}enczykowski for many useful discussions
and insights into the theoretical interpretation of the result.

\end{document}

%% file: authorlist_hepex.tex
\collab{NA48 Collaboration}
\author{A.~Lai},
\author{D.~Marras}
\address{Dipartimento di Fisica dell'Universit\`a e Sezione dell'INFN di Cagliari, \\ I-09100 Cagliari, Italy} 
\author{A.~Bevan},
\author{R.S.~Dosanjh},
\author{T.J.~Gershon},
\author{B.~Hay},
\author{G.E.~Kalmus},
\author{C.~Lazzeroni},
\author{D.J.~Munday},
\author{E.~Olaiya\thanksref{threfRAL}},
\author{M.A.~Parker},
\author{T.O.~White},
\author{S.A.~Wotton}
\address{Cavendish Laboratory, University of Cambridge, Cambridge, CB3~0HE, U.K.\thanksref{thref3}}
\thanks[thref3]{Funded by the U.K.\ Particle Physics and Astronomy Research Council}
\thanks[threfRAL]{Present address: Rutherford Appleton Laboratory, Chilton, Didcot, Oxon, OX11~0QX, U.K.}
\author{G.~Barr},
\author{G.~Bocquet},
\author{A.~Ceccucci},
\author{T.~Cuhadar-D\"onszelmann},
\author{D.~Cundy\thanksref{threfZX}},
\author{G.~D'Agostini},
\author{N.~Doble\thanksref{threfPisa}},
\author{V.~Falaleev},
\author{L.~Gatignon},
\author{A.~Gonidec},
\author{B.~Gorini},
\author{G.~Govi},
\author{P.~Grafstr\"om},
\author{W.~Kubischta},
\author{A.~Lacourt},
\author{A.~Norton},
\author{S.~Palestini},
\author{B.~Panzer-Steindel},
\author{H.~Taureg},
\author{M.~Velasco\thanksref{threfNW}},
\author{H.~Wahl\thanksref{threfHW}}
\address{CERN, CH-1211 Gen\`eve 23, Switzerland} 
\thanks[threfZX]{Present address: Istituto di Cosmogeofisica del CNR di Torino, I-10133~Torino, Italy}
\thanks[threfPisa]{Present address: Dipartimento di Fisica, Scuola Normale Superiore e Sezione dell'INFN di Pisa, I-56100~Pisa, Italy}
\thanks[threfNW]{Present address: Northwestern University, Department of Physics and Astronomy, Evanston, IL~60208, USA}
\thanks[threfHW]{Present address: Dipartimento di Fisica dell'Universit\`a e Sezione dell'INFN di Ferrara, I-44100~Ferrara, Italy}

\author{C.~Cheshkov\thanksref{threfCERN}},
\author{A.~Gaponenko},
\author{P.~Hristov\thanksref{threfCERN}},
\author{V.~Kekelidze},
\author{D.~Madigojine},
\author{N.~Molokanova},
\author{Yu.~Potrebenikov},
\author{G.~Tatishvili\thanksref{threfCM}},
\author{A.~Tkatchev},
\author{A.~Zinchenko}
\address{Joint Institute for Nuclear Research, Dubna, 141980, Russian Federation}  
\thanks[threfCERN]{Present address: EP Division, CERN, CH-1211 Gen\`eve~23, Switzerland}
\thanks[threfCM]{Present address: Carnegie Mellon University, Pittsburgh, PA~15213, USA}
\author{I.~Knowles},
\author{V.~Martin\thanksref{threfNW}},
\author{R.~Sacco\thanksref{threfSacco}},
\author{A.~Walker}
\address{Department of Physics and Astronomy, University of Edinburgh, JCMB King's Buildings, Mayfield Road, Edinburgh, EH9~3JZ, U.K.} 
\thanks[threfSacco]{Present address: Laboratoire de l'Acc\'el\'erateur Lin\'eaire, IN2P3-CNRS,Universit\'e de Paris-Sud, 91898~Orsay, France}
%
%
%
\author{M.~Contalbrigo},
\author{P.~Dalpiaz},
\author{J.~Duclos},
\author{P.L.~Frabetti\thanksref{threfFrabetti}},
\author{A.~Gianoli},
\author{M.~Martini},
\author{F.~Petrucci},
\author{M.~Savri\'e}
\address{Dipartimento di Fisica dell'Universit\`a e Sezione dell'INFN di Ferrara, \\ I-44100 Ferrara, Italy}
\thanks[threfFrabetti]{Present address: Joint Institute for Nuclear Research, Dubna, 141980, Russian Federation}
\newpage
\author{A.~Bizzeti\thanksref{threfXX}},
\author{M.~Calvetti},
\author{G.~Collazuol\thanksref{threfPisa}},
\author{G.~Graziani\thanksref{threfGG}},
\author{E.~Iacopini},
\author{M.~Lenti},
\author{F.~Martelli\thanksref{thref7}},
\author{M.~Veltri\thanksref{thref7}}
\address{Dipartimento di Fisica dell'Universit\`a e Sezione dell'INFN di Firenze, I-50125~Firenze, Italy}
\thanks[threfXX]{Dipartimento di Fisica dell'Universit\`a di Modena e Reggio Emilia, I-41100~Modena, Italy}
\thanks[threfGG]{Present address: DSM/DAPNIA - CEA Saclay, F-91191 Gif-sur-Yvette, France}
\thanks[thref7]{Istituto di Fisica dell'Universit\`a di Urbino, I-61029~Urbino, Italy}
\author{H.G.~Becker},
\author{K.~Eppard},
\author{M.~Eppard\thanksref{threfCERN}},
\author{H.~Fox\thanksref{threfNW}},
\author{A.~Kalter},
\author{K.~Kleinknecht},
\author{U.~Koch},
\author{L.~K\"opke},
\author{P.~Lopes da Silva}, 
\author{P.~Marouelli},
\author{I.~Pellmann\thanksref{threfDESY}},
\author{A.~Peters\thanksref{threfCERN}},
\author{B.~Renk},
\author{S.A.~Schmidt},
\author{V.~Sch\"onharting},
\author{Y.~Schu\'e},
\author{R.~Wanke\corauthref{cor}},
\author{A.~Winhart},
\author{M.~Wittgen\thanksref{threfSLAC}}
\address{Institut f\"ur Physik, Universit\"at Mainz, D-55099~Mainz, Germany\thanksref{thref6}}
\thanks[thref6]{Funded by the German Federal Minister for Research and Technology (BMBF) under contract 7MZ18P(4)-TP2}
\thanks[threfDESY]{Present address: DESY Hamburg, D-22607~Hamburg, Germany}
\corauth[cor]{Corresponding author.\\{\em Email address:} Rainer.Wanke@uni-mainz.de}
\thanks[threfSLAC]{Present address: SLAC, Stanford, CA~94025, USA}
\author{J.C.~Chollet},
\author{L.~Fayard},
\author{L.~Iconomidou-Fayard},
\author{J.~Ocariz},
\author{G.~Unal},
\author{I.~Wingerter-Seez}
\address{Laboratoire de l'Acc\'el\'erateur Lin\'eaire, IN2P3-CNRS,Universit\'e de Paris-Sud, 91898 Orsay, France\thanksref{threfOrsay}}
\thanks[threfOrsay]{Funded by Institut National de Physique des Particules et de Physique Nucl\'eaire (IN2P3), France}
\author{G.~Anzivino},
\author{P.~Cenci},
\author{E.~Imbergamo},
\author{P.~Lubrano},
\author{A.~Mestvirishvili},
\author{A.~Nappi},
\author{M.~Pepe},
\author{M.~Piccini}
\address{Dipartimento di Fisica dell'Universit\`a e Sezione dell'INFN di Perugia, \\ I-06100 Perugia, Italy}
\author{L.~Bertanza},
\author{R.~Carosi},
\author{R.~Casali},
\author{C.~Cerri},
\author{M.~Cirilli\thanksref{threfCERN}},
\author{F.~Costantini},
\author{R.~Fantechi},
\author{S.~Giudici},
\author{I.~Mannelli},
\author{G.~Pierazzini},
\author{M.~Sozzi}
\address{Dipartimento di Fisica, Scuola Normale Superiore e Sezione dell'INFN di Pisa, \\ I-56100~Pisa, Italy} 
\author{J.B.~Cheze},
\author{J.~Cogan},
\author{M.~De Beer},
\author{P.~Debu},
\author{A.~Formica},
\author{R.~Granier de Cassagnac},
\author{E.~Mazzucato},
\author{B.~Peyaud},
\author{R.~Turlay},
\author{B.~Vallage}
\address{DSM/DAPNIA - CEA Saclay, F-91191 Gif-sur-Yvette, France} 
%
%
%
\author{M.~Holder},
\author{A.~Maier},
\author{M.~Ziolkowski}
\address{Fachbereich Physik, Universit\"at Siegen, D-57068 Siegen, Germany\thanksref{thref8}}
\thanks[thref8]{Funded by the German Federal Minister for Research and Technology (BMBF) under contract 056SI74}
\newpage
\author{R.~Arcidiacono},
\author{C.~Biino},
\author{N.~Cartiglia},
\author{R.~Guida}, 
\author{F.~Marchetto}, 
\author{E.~Menichetti},
\author{N.~Pastrone}
\address{Dipartimento di Fisica Sperimentale dell'Universit\`a e Sezione dell'INFN di Torino, I-10125~Torino, Italy} 
\author{J.~Nassalski},
\author{E.~Rondio},
\author{M.~Szleper\thanksref{threfNW}},
\author{W.~Wislicki},
\author{S.~Wronka}
\address{Soltan Institute for Nuclear Studies, Laboratory for High Energy Physics, PL-00-681~Warsaw, Poland\thanksref{thref9}}
\thanks[thref9]{Supported by the KBN under contract SPUB-M/CERN/P03/DZ210/2000 and using computing resources of the
Interdisciplinary Center for Mathematical and Computational Modelling of the University of Warsaw.}
\author{H.~Dibon},
\author{G.~Fischer},
\author{M.~Jeitler},
\author{M.~Markytan},
\author{I.~Mikulec},
\author{G.~Neuhofer},
\author{M.~Pernicka},
\author{A.~Taurok},
\author{L.~Widhalm}
\address{\"Osterreichische Akademie der Wissenschaften, Institut f\"ur Hochenergiephysik, A-1050~Wien, Austria\thanksref{thref10}}
\thanks[thref10]{Funded by the Federal Ministry od Science and Transportation under the contract GZ~616.360/2-IV GZ 616.363/2-VIII, 
and by the Austrian Science Foundation under contract P08929-PHY.}